\documentclass[12pt]{article}
\usepackage{epsfig,amsfonts,amssymb}
\usepackage{hyperref}
\usepackage{cite}
\input epsf.sty
\def\ZZZ{{\hbox{ Z\kern-1.6mm Z}}}
\def\zzz{{\hbox{z\kern-1mm z}}}

\newcommand{\vt}{\vartheta}

\def\RRR{{\hbox{ R\kern-2.4mm R}}}

\newcommand{\CC}{{\cal C}}

\newcommand{\LL}{{\cal L}}

\newcommand{\wt}{\widetilde}

\newcommand{\RR}{{\cal R}}
\newcommand{\NN}{{\cal N}}

\newcommand{\be}{\begin{equation}}
\newcommand{\ee}{\end{equation}}
\newcommand{\ben}{\begin{eqnarray}\displaystyle}
\newcommand{\een}{\end{eqnarray}}

\newcommand{\bea}[1]{\begin{eqnarray}\label{#1} }
\newcommand{\eea}{\end{eqnarray}}

\newcommand{\refb}[1]{(\ref{#1})}
\newcommand{\p}{\partial}
\newcommand{\sectiono}[1]{\section{#1}\setcounter{equation}{0}}

\def\one{{\hbox{ 1\kern-.8mm l}}}
\def\zero{{\hbox{ 0\kern-1.5mm 0}}}

\begin{document}

\baselineskip 24pt

\begin{center}

{\Large \bf How Do Black Holes Predict the
Sign of the Fourier Coefficients
of Siegel Modular Forms?}

\end{center}

\vskip .6cm
\medskip

\vspace*{4.0ex}

\baselineskip=18pt

\centerline{\large \rm Ashoke Sen}

\vspace*{4.0ex}
\centerline{\large \it Harish-Chandra Research Institute}
\centerline{\large \it  Chhatnag Road, Jhusi,
Allahabad 211019, India}
\centerline{and}
\vspace*{3.0ex}
\centerline{\large \it 
LPTHE, Universite Pierre et Marie Curie, Paris 6}
\centerline{\large \it 
4 Place Jussieu,  75252 Paris Cedex 05, France}

\vspace*{1.0ex}
\centerline{E-mail:  sen 
at hri.res.in, ashokesen1999 at gmail.com}

\vspace*{5.0ex}

\centerline{\bf Abstract} \bigskip

Single centered supersymmetric black holes in four
dimensions have spherically symmetric horizon and hence
carry zero angular momentum. This leads to a specific sign
of the helicity trace index associated with these black holes.
Since the latter are given by the Fourier
expansion
coefficients of appropriate meromorphic 
modular forms of $Sp(2,\ZZZ)$ or its subgroup,
we are led to a specific prediction for the signs of a
subset of these Fourier coefficients which represent
contributions from single centered black holes only.
We explicitly test these predictions for 
the modular forms which compute
the index of quarter BPS black holes in heterotic string theory
on $T^6$, as well as in $\ZZZ_N$ CHL models for $N=2,3,5,7$.

\vfill \eject

\baselineskip=18pt

\tableofcontents

\renewcommand{\theequation}{\thesection.\arabic{equation}}

\sectiono{Introduction} \label{s00}

Classical single centered black holes 
in four dimensions are 
spherically
symmetric and hence carry zero angular momentum.
Since the black hole breaks part of the supersymmetry
of the theory, supersymmetric excitations around the black
hole include a set of fermion zero modes, and hence
quantization of these fermion zero modes impart certain
angular momentum on the black hole. However these
fermion zero modes live outside the horizon, and
the horizon of the black hole continues to 
remain spherically symmetric
as a consequence of supersymmetry.
Given the folklore that black holes describe average
properties of an ensemble one might tend to conclude
that spherical symmetry implies zero average
angular momentum
carried by the black hole, -- with the individual members
of the ensemble carrying different angular momentum.
 However
using $AdS_2/CFT_1$ correspondence it has been argued 
in \cite{0903.1477} that
a spherically symmetric horizon implies that the black hole
represents a microcanonical ensemble of states {\it all of
which carry zero angular momentum}.
Thus the only source of angular momentum carried by
the black hole is from the fermion zero modes associated
with broken supersymmetry. This in turn implies that
the helicity trace index of the black hole, 
defined as\cite{9611205,9708062}
\be \label{e1}
B_{2n} = {1\over (2n)!} \, Tr((-1)^F (2h)^{2n})\, ,
\ee
is given by $(-1)^n\, d_{hor}$ where $d_{hor}$ is
the degeneracy of the ensemble represented by the
horizon of the black hole.
Here $F$ denotes fermion number,
$h$ denotes the third component of the angular momentum
carried by the black hole in its rest frame,  the trace
is taken over all states carrying a given set of charges, and
$4n$ is the number of supersymmetries broken by the
black hole, which is equal to the number of
fermion zero modes on the black hole.  The result quoted
above follows from the fact that quantization of each pair
of fermion zero modes produces a pair of states with
$h=\pm{1\over 4}$ and hence $Tr\{(-1)^F (2h)\} 
= Tr\{e^{2\pi i h} (2h)\}=i$.
Thus $2n$ pairs of fermion zero modes will give a 
contribution to $B_{2n}$ of the form $i^{2n}=(-1)^n$.
The factor of $1/(2n)!$ in the definition of $B_{2n}$ cancels
against a combinatoric factor that appears when we write
$2h$ as the sum of the contribution from individual pairs
of fermion zero modes and carry out a binomial expansion 
of $(2h)^{2n}$,
picking up
the term that contains one factor of $2h$ for each pair of
fermion zero modes.
Once the trace of the fermion zero modes has 
been performed, we just need to evaluate $Tr(-1)^F$
over the rest of the degrees of freedom, and the horizon
contribution to this is the same as the degeneracy $d_{hor}$
since
$(-1)^F=1$ for all the states represented by the 
horizon\cite{0903.1477,appear}.

We shall focus on quarter BPS black holes in $\NN=4$
supersymmetric string theories which break 12 out of
16 supersymmetries and hence the relevant index is
$B_6$. The analysis given above predicts that 
$B_6=-d_{hor}$.
$d_{hor}$ can be calculated in 
principle using quantum entropy function 
formalism\cite{0809.3304}, but for our argument
the only relevant fact about $d_{hor}$ will be that
being a degeneracy it must be positive. This in turn
implies that 
$B_6$ must be negative\cite{0903.1477}. 

There are several effects which could potentially destroy
this prediction.
\begin{enumerate}
\item For given set of charges and a given point in the moduli
space of the theory the index may receive contribution not
only from single centered black holes but also
multi-centered black holes. Since multi-centered black
holes can carry angular momentum from the fields
living outside the black hole 
horizons\cite{0010222,0101135,0206072,0304094,0702146} 
there is no
longer any guarantee that the contribution to $B_6$ from
these black holes will be negative. 
This problem can however
be easily avoided by working in a 
chamber of the moduli space
bounded by the walls of marginal stability 
that contains the
attractor point. In this chamber only single 
centered black holes
contribute to the index\cite{0706.2363,0806.2337} 
and our prediction 
for the sign of $B_6$
holds.\footnote{In a subspace of the moduli space where
a multi-centered configuration can be embedded in an
$\NN=2$ supersymmetric string theory, there  exist
a family of solutions known as
scaling solutions\cite{0702146} 
which continue to exist
even at the attractor point. At a generic point in the moduli
space of $\NN=4$ supersymmetric string theory
we do not expect these solutions to exist since
they cannot be embedded in an $\NN=2$ supersymmetric
theory where they have been constructed\cite{0903.2481}. 
}
We shall refer to this chamber of the moduli space as
the attractor chamber.
\item Another source of breakdown of our argument is
the possible existence of additional supersymmetry
preserving fermionic excitations outside
the horizon (hair modes\cite{0901.0359,0907.0593}) 
besides the fermionic zero modes associated
with broken supersymmetry. Quantization of these
modes would give both $(-1)^F$ odd and $(-1)^F$ even 
states, and this could turn
a positive contribution to $Tr(-1)^F$ from the horizon
into a negaive contribution.
This   can in principle be avoided by going to a
duality frame in which 
all the charges carried by the black hole correspond
to some kind of brane charges rather than momenta
along compact circles. Since the hair modes described
in \cite{0901.0359,0907.0593} come from 
excitations carrying momentum along
some compact directions, this type of hair modes can be
avoided if the black hole does not carry any net
momentum along any of the internal directions.
\item The final source of breakdown of our argument
arises from the possibility that in a given charge sector the
contribution to the index could come
from horizonless smooth solutions
besides the black hole. Indeed a wide class of
smooth solutions have been constructed in supergravity
theories (see {\it e.g.}
\cite{1006.3497} and references therein). 
If such solutions exist then their contribution to the index
must be added to that from the black hole\cite{0908.3402}
and this could
potentially change a negative $B_6$ of the black hole into
a positive value. 
However it is not obvious that these smooth solutions, even if
they exist at a generic point in the moduli space, would
contribute to the index. Typically in $\NN=4$ supersmmetric
theories it is  difficult to construct classical 
solutions which contribute to the index except in 
very special cases. As an example one can 
mention multi-centered black holes or two centered black
holes at least one of whose centers is quarter BPS.
These exist as supersymmetric classical solutions
in a subspace of the moduli space of the theory where the
solution can be embedded in an $\NN=2$ supersymmetric
theory. But their contribution to $B_6$ must vanish as can
be seen from the fact that one can find a continuous path
in the moduli space of $\NN=4$ supersymmetric string
theory that does not hit any wall of marginal stability and yet
reaches a point where these solutions do not 
exist\cite{0707.1563}.
Physically the vanishing of the index can be understood
as due to the difficulty in aligning the supersymmetries
of different parts of the solution\cite{0803.3857,0903.2481}.
The essential point is that since a quarter BPS solution
breaks 12 out of 16 supersymmetries, each part of the
solution aligns its 4 unbroken supersymmetries 
in a certain  way in the space of 16 supersymmetries.
In order that the full solution is supersymmetric the
supersymmetries of different parts must be compatible,
\i.e.\ the four unbroken supersymmetries of different parts
must align appropriately inside the 
space of 16 supersymmetries. This is a stronger requirement
in $\NN=4$ supersymmetric theory than in 
$\NN=2$ supersymmetric theory since in the latter case
the full theory has 8 supersymmetries and hence the
four unbroken supersymmetries of different parts need to be
aligned inside the space of 8 
supersymmetries.\footnote{This argument can be 
made more precise in terms of
alignment of central charges, --  the central charge in an
$\NN=2$ supersymmetric theory is a two dimensional real vector
while in an $\NN=4$ supersymmetric theory it is a six dimensional
real vector. Clearly it is easier to align several two dimensional
vectors compared to several six dimensional 
vectors\cite{0903.2481}.}
Due to this reason having a classical solution 
that contributes
to the $B_6$ index in $\NN=4$ supersymmetric theories is
more unlikey than in its $\NN=2$ counterpart, and we shall
assume that such solutions do not exist for the range of charges
for which a single centered black hole solution exists.
\end{enumerate}

So 
we shall proceed with the assumption that there
exists some duality frame in which only single centered
black hole solution -- whose only hair are the fermion
zero modes associated with broken supersymmetry --
contributes to $B_6$
in the attractor chamber.
As a result $B_6$ must be negative.
We shall now try to test this prediction using known
microscopic results.

\sectiono{The result for the index} \label{stwo}

The index $B_6$ has been calculated in a wide class of
$\NN=4$ supersymmetric string theories for a wide class
of charges\cite{9607026,0412287,0505094,
0506249,0508174,0510147,0602254,0603066,0605210,
0607155,0609109,0612011,0705.1433,0708.1270,0802.0544,
0802.1556,0803.2692,0911.1563,1002.3857,1006.3472}
(see \cite{1008.3801} for a recent survey of the results). 
It is convenient to label the charges carried
by the state by a pair of (electric, magnetic) charge vectors
$(Q,P)$ in a frame where we represent the theory as (an
orbifold of) heterotic string theory compactified on
$T^6$. We shall denote by $Q^2$, $P^2$ and $Q\cdot P$ the
continuous T-duality invariant inner products of $Q$ and $P$ in
this duality frame. Then in the $\ZZZ_N$ CHL 
models\cite{9505054,9506048},
obtained by taking an appropriate $\ZZZ_N$ quotient of
heterotic string theory on $T^6$, the result for $B_6$ 
takes the form:\footnote{Note that this result does not hold for
all dyons but a subset of dyons belonging to specific duality
orbits in these theories.}
\be\label{egg1int}
B_6(\vec Q,\vec P) = {1\over N}\, (-1)^{Q\cdot P}\,
\int _\CC d \rho \, 
d \sigma \,
d  v \, e^{-\pi i ( N  \rho Q^2
+   \sigma P^2/N +2  v Q\cdot P)}\, {1
\over  \wt\Phi(  \rho,  \sigma,   v)}\, ,
\ee
where for any given $N$,
$\wt\Phi(\rho,\sigma,v)$ is a known
function, transforming
as a modular form of certain weight under a subgroup
of $Sp(2,\ZZZ)$\cite{igusa1,igusa2,borcherds,9504006,ibu1,ibu2,ibu3,
eichler,skor,rama}, and $\CC$ is a three real 
dimensional subspace of the
three complex dimensional space labelled by $( \rho=\rho_1+i
\rho_2,\sigma=\sigma_1+i\sigma_2, v=v_1+i v_2)$. 
Eq.\refb{egg1int} encompasses the $N=1$ 
case that describes
heterotic string theory on $T^6$.
The contour $\CC$ takes the form:
\ben \label{ep2kk}
 \rho_2=M_1, \quad   \sigma_2 = M_2, \quad
   v_2 = -M_3, \nonumber \\
 0\le   \rho_1\le 1, \quad
0\le   \sigma_1\le N, \quad 0\le    v_1\le 1\, ,
\een
where $M_1$, $M_2$ and $M_3$ are large but fixed real
numbers. The choice of $(M_1,M_2,M_3)$ is governed by
the chamber in the moduli space in which we want to compute
the index\cite{0702141,0702150} 
-- there being a one to one correspondence between
the chambers in the moduli space separated by walls
of marginal stability and the domains in the $(M_1,M_2,M_3)$
space separated by poles. The jump in $B_6$ across a wall
of marginal stability is given by the residue of the integrand
at the pole that separates the corresponding domains, and is
in accordance with the wall crossing 
formula\cite{0705.3874,0706.2363}.

\begin{figure}
\leavevmode
\begin{center}
\hbox{
\epsfysize=3cm \epsfbox{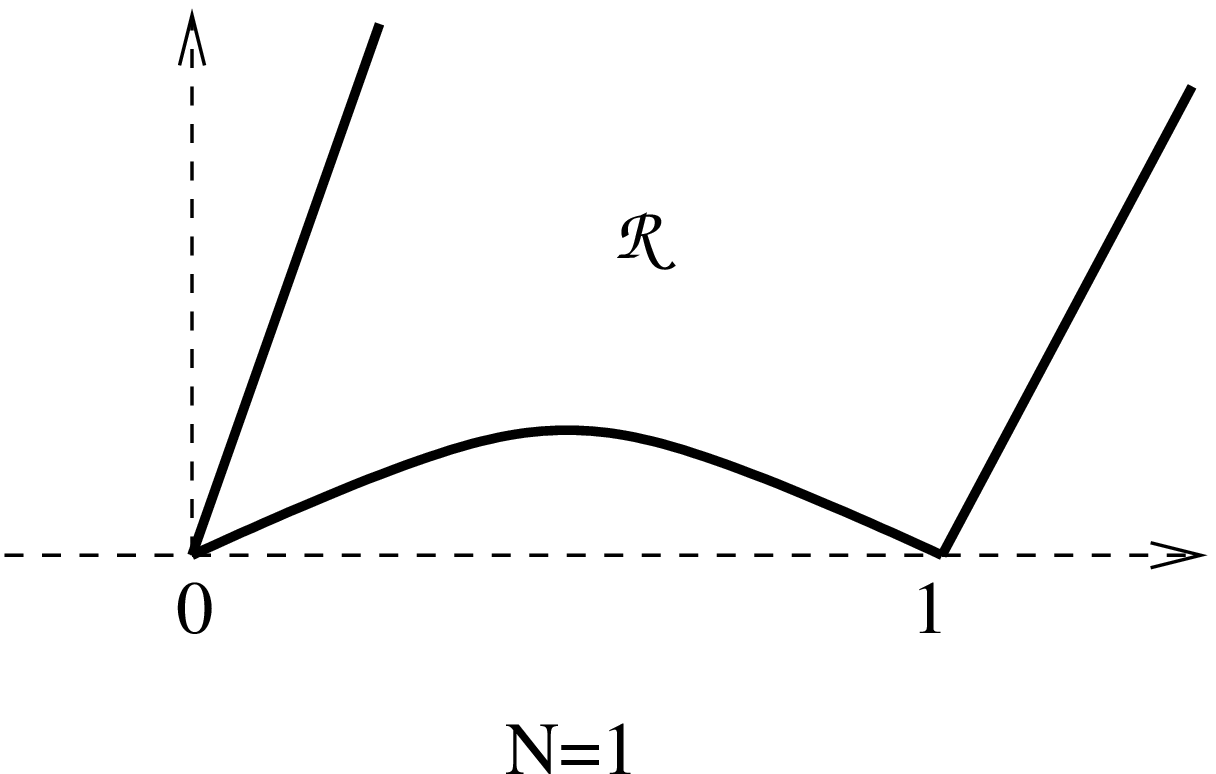}
\epsfysize=3cm \epsfbox{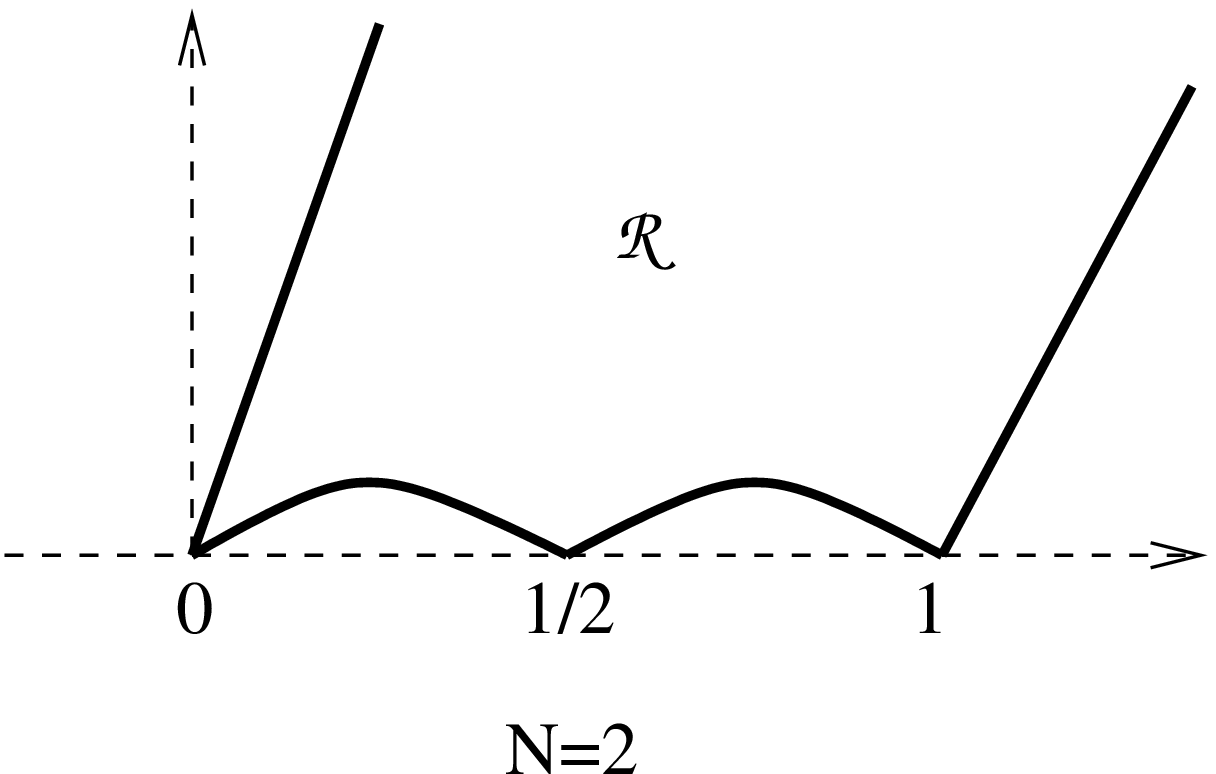}
\epsfysize=3cm \epsfbox{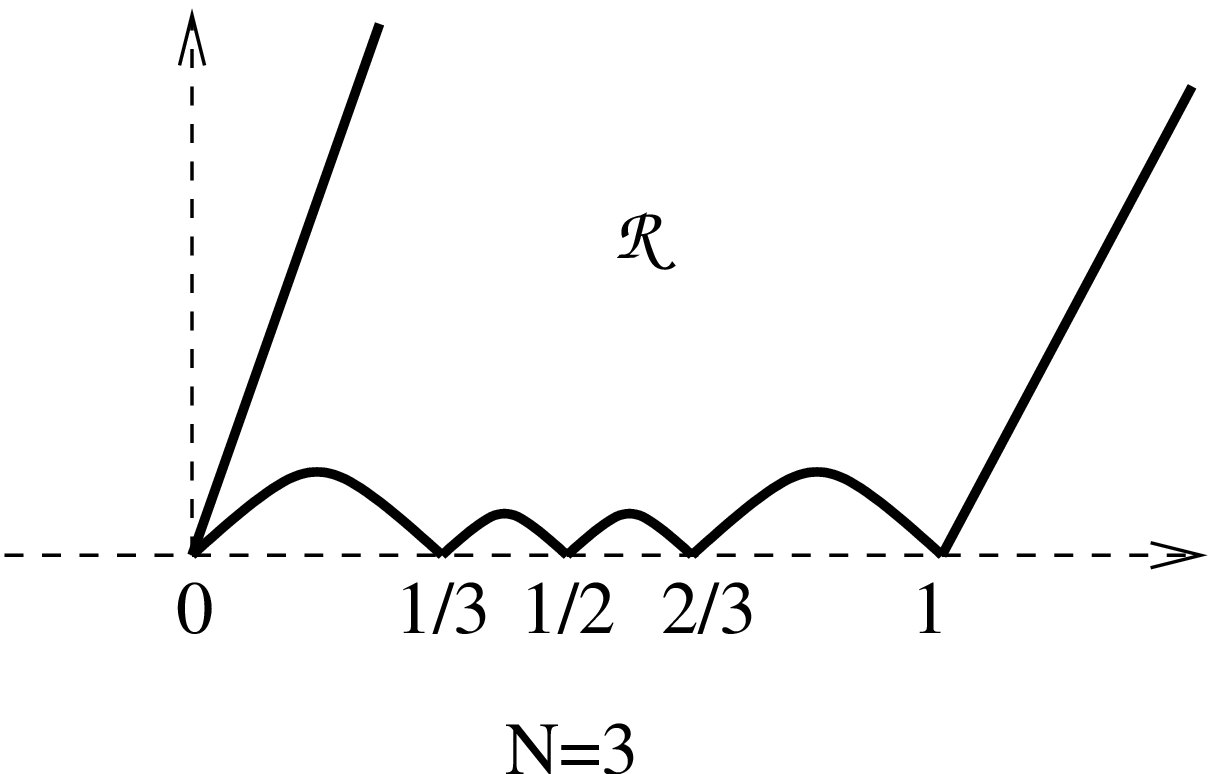}
}
\end{center}
\caption{A schematic diagram representing the chamber $\RR$
in the upper half $\tau$ plane, bounded by the walls of
marginal stability,  for $\ZZZ_N$
orbifolds of heterotic string theory on $T^6$ for
$N=1,2,3$. 
The shapes of the circles and
the slopes of the straight lines bordering the chamber depend
on the charges and other asymptotic moduli, but the vertices
are universal.} \label{f1}
\end{figure}

For large charges the contribution from single centered 
black holes is the dominant contribution 
in all chambers\cite{0707.1563,0903.2481}
and
hence the argument presented in
\S\ref{s00} will imply that $B_6$
is negative in all the chambers\cite{0903.1477}. 
This has been explicitly
verified by analyzing the behaviour of \refb{egg1int} for
large charges\cite{0708.1270}. 
Our goal is to verify the prediction for
the sign of $B_6$ for finite charges, and for this we must work
in the attractor chamber. There are several approaches we can
follow. For a given $(Q,P)$ we can determine the values of
$(M_1, M_2, M_3)$ when the moduli are at the attractor point,
-- a general algorithm for finding this has been given in
\cite{0706.2363}. One can also try 
to first define a generating function for
single centered black holes starting from \refb{egg1int} and
use it to extract the $B_6$ indices for single centered 
black holes\cite{atish}.
We shall follow a third approach which we find most practical.
Fig.~\ref{f1} shows the shapes of the some of the
walls of marginal stability
in the heterotic axion-dilaton moduli space labelled by the complex
field $\tau$ taking values in the upper half plane, 
for fixed values of the other moduli\cite{0702141}. 
We shall denote by
$\RR$ a specific chamber that lies just to the right of the
wall that connects 0 to $i\infty$ in the $\tau$ plane,
determine the constraints on the charges that makes the
attractor point lie inside the chamber $\RR$, and verify that
$B_6$ in $\RR$ is negative for all these charges. 
Since
for heterotic string theory on $T^6$, and for $\ZZZ_2$
and $\ZZZ_3$ CHL models, every chamber can be mapped
to $\RR$ by an S-duality 
transformation\cite{0702141}, this would
prove that for all single centered black holes $B_6$ is
negative, provided the charges carried by the black hole fall on
the duality orbit for which \refb{egg1int} holds.
Whether there exist duality transformations mapping every chamber
to $\RR$ is not known for the $\ZZZ_5$ and $\ZZZ_7$
CHL models. Nevertheless the negativity of $B_6$ for single
centered dyons in $\RR$ is a
necessary condition which can be tested even in these models.

The choice
of $(M_1,M_2,M_3)$ corresponding to the chamber $\RR$
is\cite{0609109,0702141}:
\be \label{emam2m3}
M_1, M_2 >> 0, \quad M_3 << 0, \quad |M_3| << M_1, M_2\, .
\ee
In practical terms this means that to extract $B_6$ in
this chamber we first expand $1/\wt\Phi$ in powers of 
$e^{2\pi i\rho}$ and $e^{2\pi i\sigma}$ and then expand
each term in this expansion in powers of $e^{-2\pi i v}$.
This is best done using the product representation of
$\wt\Phi$. 
For the $\ZZZ_N$ CHL models with $N=1,2,3,5,7$
we 
have\cite{0602254,0605210,0609109}
\ben \label{edefwtphi}
&&  \wt \Phi(  \rho,  \sigma,  v )^{-1} =
e^{-2\pi i ( \rho +   \sigma / N +   v)} \nonumber \\
&& \qquad \times \prod_{b=0}^1\, 
 \prod_{r=0}^{N-1}
\prod_{k\in \zzz+{r\over N},l\in\zzz,j\in 2\zzz+b
\atop k,l\ge 0, j<0 \, {\rm for}
\, k=l=0}
\left( 1 - \exp\left(2\pi i ( k  \sigma   +  l  \rho +  j  v)
\right)\right)^{-
\sum_{s=0}^{N-1} e^{-2\pi i sl/N } c^{(r,s)}_b(4kl - j^2)} \, ,
\nonumber \\
\eea
where the coefficients 
$c_b^{(r,s)}(u)$ are defined as follows\cite{0602254}.
First we define\footnote{A different but equivalent description of
the functions $F^{(r,s)}$ can be derived from the general result of
\cite{1002.3857}.}
\bea{fifthn}
F^{(0,0)}(\tau, z) &=& {8\over N} A(\tau, z)\, ,
\nonumber \\
F^{(0,s)}(\tau, z) &=& {8\over N(N+1)} \, A(\tau, z) -{2\over N+1}
\, B(\tau, z) \, E_N(\tau) \qquad \hbox{for $1\le s\le (N-1)$}
\, , \nonumber \\
F^{(r,rk)}(\tau, z) &=& {8\over N(N+1)} \, A(\tau, z)
+ {2\over N(N+1)} \, E_N\left({\tau+k\over N}\right)\, B(\tau, z)\, ,
\nonumber \\
&& \qquad \qquad \qquad \qquad 
\qquad \hbox{for $1\le r \le (N-1)$, $0\le k\le (N-1)$}
\, ,\nonumber \\
\eea
where
\be\label{efirstn}
 A(\tau, z) =  \left[ {\vartheta_2(\tau,z)^2
\over \vartheta_2(\tau,0)^2} +
{\vartheta_3(\tau,z)^2\over \vartheta_3(\tau,0)^2}
+ {\vartheta_4(\tau,z)^2\over \vartheta_4(\tau,0)^2}\right]\, ,
\ee
\be\label{secondn}
B(\tau, z) = \eta(\tau)^{-6} \vartheta_1(\tau, z)^2\, ,
\ee
and
\be\label{thirdn}
E_N(\tau) = {12 i\over \pi(N-1)} \, \p_\tau \left[ \ln\eta(\tau)
-\ln\eta(N\tau)\right]= 1 + {24\over N-1} \, \sum_{n_1,n_2\ge 1\atop
n_1 \ne 0 \,  {\rm \small mod} 
\, N} n_1 e^{2\pi i n_1 n_2 \tau}\, .
\ee
Then $c_b^{(r,s)}(u)$ is defined via the expansion:
\be \label{ecomp}
F^{(r,s)}(\tau,z)\equiv
\sum_{b=0}^1\sum_{j\in2\zzz+b, n\in \zzz/N} 
c^{(r,s)}_b(4n -j^2)
e^{2\pi i n\tau + 2\pi i jz}\, .
\ee
For terms in \refb{edefwtphi}
with either $l$ or $k$ non-zero, the procedure
of expansion
is straightforward; we simply expand
the $\left( 1 - e^{2\pi i ( k  \sigma   +  l  \rho +  j  v)
}\right)^{-
\sum_{s=0}^{N-1} e^{-2\pi i sl/N } c^{(r,s)}_b(4kl - j^2)}$
term in a power series in 
$e^{2\pi i ( k  \sigma   +  l  \rho +  j  v)}$.
Special care needs to be taken for the the $k=l=0$ term
which, together with the $e^{-2\pi i v}$ factor in the front,
is given by $e^{-2\pi i v} / (1 - e^{-2\pi i v})^2$. The
contour prescripton for chamber $\RR$, corresponding to
the choice of $M_i$ given in \refb{emam2m3},
requires us to expand this factor in powers of $e^{-2\pi i v}$.
This gives a completely well defined prescription for
expanding $1/\wt\Phi$ and computing $B_6$ in the chamber 
$\RR$.

\sectiono{Kinematic constraints on the charges} \label{sthree}

Now that we have described the algorithm for calculating
$B_6$ in the chamber $\RR$, the next question we need
to ask is: for which charges $(Q,P)$ the attractor point
in the moduli space lies inside $\RR$? Once we 
determine these charges, our previous argument will
tell us that $B_6(Q,P)$ for these charges, computed
inside the chamber $\RR$, must be negative. There are
various approaches to answer this question,  
we shall describe one of them.

We begin with the $N=1$ model, \i.e.\ heterotic string
theory on $T^6$.
First consider the wall that connects 0 to
$i\infty$. For reasons which will become clear soon, we
shall assign an orientation to this line which we take
to be directed away from 0 and towards the point at
$i\infty$. A necessary condition that the attractor point lies
inside $\RR$ is that it lies to the right of the wall
going from 0 to $i\infty$. Now if we denote by $M$
the symmetric $SO(6,22)$ matrix valued 
moduli of the string theory ($SO(6,22)$ will be
replaced by $SO(6,r)$ for some
other integer $r$ for CHL models), 
by $L$ the $O(6,22)$
invariant matrix of signature $(+^6-^{22})$, and
by
\be \label{edefqrpr}
Q_R = {1\over 2} (M+L)Q, \qquad 
P_R = {1\over 2} (M+L)P,
\ee
then at the attractor point
\be \label{eattr}
Q_R^2 = Q^2, \quad P_R^2 = P^2, \quad Q_R . P_R
= Q . P\, ,
\ee
\be \label{eattr2}
\tau_1 = {Q . P\over P^2}, \quad
\tau_2 = {\sqrt{Q^2 P^2 - (Q . P)^2}\over P^2}\, .
\ee
On the other hand the wall of marginal stability
joining 0 and $i\infty$ is described by the 
equation\cite{0702141}
\be \label{ewall}
\tau_1 +  {Q_R . P_R\over 
\sqrt{Q_R^2 P_R^2 - (Q_R. P_R)^2}}\tau_2 =0\, .
\ee
If we choose $M$ and $\tau_2$ to be at their
attractor values given in \refb{eattr}, \refb{eattr2} then
the value of $\tau_1$ computed from \refb{ewall}
is given by $-Q . P/P^2$. Thus in order that the
attractor point lies to the right of this wall, we need
$\tau_1$ given in \refb{eattr2} to be larger than
$-Q . P/P^2$, \i.e.\ have
$Q . P/P^2\ge 0$.\footnote{For 
$Q.P=0$ a two centered black hole carrying charges
$(Q,0)$ and $(0,P)$ may exist, but its contribution to the
index, being proportional to $Q.P$, vanishes.
For this reason we have used $\ge$ instead of $>$.}
Since we shall always consider the range in
which $Q^2,P^2>0, (Q.P)^2<Q^2P^2$ 
(non-singular supersymmetric black holes exist
only in this range) we must have
\be \label{ewx}
Q . P\ge 0\, .
\ee

Since the equations for the other walls of $\RR$ are also 
known\cite{0702141,0707.3035}
we can use similar method to determine the condition
on the charges which will ensure that the attractor point
lies inside $\RR$. But we shall now describe  a simpler
method for determining this using S-duality transformation
that acts simultaneously on the charges and 
the $\tau$-moduli
as
\be \label{esd}
\tau'= {a\tau+b\over c\tau+d}, \quad \pmatrix{Q'\cr P'}
= \pmatrix{a & b\cr c & d}\pmatrix{Q\cr P}\, ,
\quad \pmatrix{a & b\cr c & d}\in SL(2,\ZZZ)\, .
\ee
If we consider the wall from 0 to 1 in the
$\tau$-plane then the $SL(2,\ZZZ)$ transformation
by $\pmatrix{1 & 0\cr -1 & 1}$ maps it to a wall
from 0 to $i\infty$ in the $\tau'$ plane. Now in order that
the attractor point corresponding to the charge $(Q,P)$
in the $\tau$ plane lies inside $\RR$ it must lie to the {\it left}
of the wall from 0 to 1. Thus in the $\tau'$ plane the
attractor point for $(Q',P')$ must lie to the left of the
wall from 0 to $i\infty$. From our previous analysis this
requires $Q' . P'\le 0$. Now from \refb{esd} we have
$(Q'=Q, P'=P-Q)$ and hence the condition $Q' . P'\le 0$
translates to 
$Q . P\le Q^2$.
Similarly mapping the wall from 1 to $i\infty$ to the wall
from 0 to $i\infty$ by the transformation
$\tau'=\tau-1$ we get the third condition 
$Q . P\le P^2$. Together with these three conditions
we must add the conditions $Q^2, P^2,
\{Q^2 P^2 - (Q . P)^2\} > 0$
since classical black hole solutions with non-singular event
horizon exists only when this condition is satisfied.
Thus we would conclude the for heterotic string theory
on $T^6$ the $B_6$ index in $\RR$
must be negative when all of
the following conditions are satisfied:
\be \label{eb6neg}
Q . P\ge 0, \quad Q . P\le Q^2, \quad 
Q . P\le P^2, \quad Q^2, P^2,
\{Q^2 P^2 - (Q . P)^2\} > 0\, .
\ee

Similar analysis can be performed for the 
CHL models obtained by taking the $\ZZZ_N$
orbifold of heterotic string theory on $T^6$. Let us
first consider the case of $N=2$ for which
the region $\RR$ is bounded by four walls shown
in Fig.\ref{f1}. In this case the
S-duality group is $\Gamma_1(2)$. 
As before the wall connecting 0 and
$i\infty$ gives the condition $Q . P\ge 0$. Now the
other walls from 0 to 1/2, 1 to 1/2 and 1 to $i\infty$
can all be mapped to the wall from 0 to $i\infty$ with
the help of $\Gamma_1(2)$ 
transfrmations\footnote{Even though 
we have used $\Gamma_1(N)$ transformations
to map the walls bordering $\RR$ to the wall connecting 0 and
$i\infty$, this is not necessary. The walls are results of kinematical
constraints and transform covariantly under any $SL(2,R)$
transformation. Thus given a wall 
connecting a point $\tau=a$ to $\tau=b$
with $b>a$, we can
use the $SL(2,R)$ transformaton $(b-a)^{-1/2}
\pmatrix{1 & -a\cr -1 & b}$
to map it to the wall connecting 0 and
$i\infty$. The constraint on $(Q,P)$ in order that the point
lies above the wall connecting $p$ and $q$ now translates to
$(Q - aP).(bP-Q)\le 0$.
}
\be \label{egam1n}
\pmatrix{1 & 0\cr -2 & 1}, \quad \pmatrix{-1 & 1\cr -2 & 1},
\quad \pmatrix{1 & -1\cr 0 & 1}\, ,
\ee
respectively. This can be used to derive the following 
conditions on $(Q,P)$ for the attractor point to lie inside
the region $\RR$:
\be \label{egam2n}
Q . P\ge 0, \quad Q . P\le 2\, Q^2, \quad 
Q . P\le P^2, \quad
3 \, Q . P \le 2 \, Q^2 + P^2,
\quad Q^2, P^2,
\{Q^2 P^2 - (Q . P)^2\} > 0\, .
\ee
The same analysis can be repeated for $N=3$. The
walls
from 0 to 1/3, 1/2 to 1/3, 1/2 to 2/3, 1 to 2/3 and 1 to
$i\infty$ are mapped to the wall from 0 to $i\infty$ via
the $\Gamma_1(3)$ transformations
\be \label{egan3n}
\pmatrix{1 & 0\cr -3 & 1}, \quad 
\pmatrix{-2 & 1\cr -3 & 1}, \quad 
\pmatrix{-2 & 1\cr 3 & -2}, \quad 
\pmatrix{1 & -1\cr 3 & -2}, \quad 
\pmatrix{1 & -1\cr 1 & 0}\, .
\ee
The conditions on $(Q,P)$ for the attractor point to lie
inside the region $\RR$ is
\ben \label{econd}
&& Q . P\ge 0, \quad Q . P\le 3\, Q^2, \quad 
Q . P\le P^2, \quad
5\, Q . P \le 6 \, Q^2 + P^2, \quad
5 \, Q . P \le 3 \, Q^2 + 2\, P^2, \nonumber \\ &&
7 \, Q . P \le 6 \, Q^2 + 2\, P^2,
\quad Q^2, P^2,
\{Q^2 P^2 - (Q . P)^2\} > 0\, .
\een

To summarize, our argument of \S\ref{s00} 
predicts that $B_6$ computed in the region $\RR$ must
be negative for $(Q^2,P^2,Q.P)$ satisfying the
constraints \refb{eb6neg} for heterotic string theory on
$T^6$, the constraints \refb{egam2n} for the $\ZZZ_2$
CHL model, and the constraints \refb{econd} for the
$\ZZZ_3$ CHL model. Since various mathematical
properties of $\wt\Phi$ have been analyzed in
\cite{0806.2337,0807.4451,0809.4258,0904.4253,
0907.1410}, 
it will be interesting to see if these predictions
follow from these properties.

For $N>3$ the number of walls bordering $\RR$ becomes 
infinite\cite{0702141} and so there are 
infinite number of constraints. 
The wall from 0 to $i\infty$ still gives the constraint $Q.P\ge 0$.
Thus if we can show, for the range of $(Q^2,P^2)$ for
which we carry out the analysis, that $B_6$ is negative 
for all $Q.P$ satisfying
\be \label{enes}
Q.P\ge 0, \qquad (Q.P)^2 < Q^2 P^2, \qquad Q^2, P^2 > 0,
\ee
then it will imply that $B_6$ is negative for single centered
dyons in this range of charges.
Note that this test is sufficient but not necessary; if we
find a positive $B_6$ value for some charges 
satisfying \refb{enes} then it
may still be consistent with our result if 
the charges fail to satisfy any of the other conditions
associated with the other walls of $\RR$.

\begin{table} {\small
\begin{center}\def\st{\vrule height 3ex width 0ex}
\begin{tabular}{|l|l|l|l|l|l|l|l|l|l|l|} \hline 
$(Q^2,P^2){\backslash} Q.P$ &  -2 
& 0 & 1 & 2 & 3 & 4\st\\[1ex] \hline \hline
(2,2) &   -209304 &  {\bf 50064}
&  {\bf 25353} &  648 & 327 & 0 \st\\[1ex] \hline
(2,4) &  -2023536  & {\bf 1127472}
&  {\bf 561576} & {\bf 50064} & 8376 & -648 \st\\[1ex] \hline
(4,4) & -16620544  &  {\bf 32861184} & {\bf 18458000} &  {\bf 3859456}
&  {\bf 561576} & 12800 \st\\[1ex] \hline
(2,6) &  -15493728 &  {\bf 16491600}
&  {\bf 8533821} & {\bf 1127472} & 130329 & -15600 \st\\[1ex] \hline
(4,6) & -53249700 &  {\bf 632078672} & {\bf 392427528}   & 
{\bf 110910300}  &  
{\bf 18458000} 
&  {\bf 1127472} \st\\[1ex] \hline
(6,6) & 2857656828  &  {\bf 16193130552} & {\bf 11232685725}  & 
{\bf 4173501828}
&  {\bf 920577636} & {\bf 110910300}  \st\\[1ex] \hline
 \hline 
\end{tabular}
\caption{Some results for $-B_6$ in heterotic string
theory on $T^6$ for different values of $Q^2$, $P^2$ and
$Q . P$. The boldfaced entries are for charges
which satisfy the constraints \refb{eb6neg}.
We have given the results only for $Q^2\le P^2$ since
the results are symmetric under $Q^2\leftrightarrow P^2$.
Note that some of the entries are the same; this is
a consequence of a $\ZZZ_3$ subgroup of S-duality
transformation $\tau\to 1 -\tau^{-1}$ which maps $\RR$
to $\RR$ but changes the charges as $(Q^2,P^2,Q\cdot P)
\to (P^2+Q^2-2Q.P, Q^2, Q^2-Q.P)$.
} \label{t1}
\end{center} }
\end{table}

\begin{table} {\small
\begin{center}\def\st{\vrule height 3ex width 0ex}
\begin{tabular}{|l|l|l|l|l|l|l|l|l|l|l|} \hline 
$(Q^2,P^2)\backslash Q.P$ &  -2 
& 0 & 1 & 2 & 3 & 4\st\\[1ex] \hline \hline
(1,2) & -5410   &  {\bf 2164}
&  {\bf 360} &  -2 & 0 & 0 \st\\[1ex] \hline
(1,4) & - 26464    & {\bf 18944}
&  {\bf 4352} & {160} & 0 & 0 \st\\[1ex] \hline
(2,4) & -124160  &  {\bf  198144} & {\bf  67008} &  {\bf 6912}
&  {64} &  0 \st\\[1ex] \hline
(1,6) & -114524  &  {\bf  125860}
&  {\bf  36024} & {\bf 2164} & 52  & 0  \st\\[1ex] \hline
(2,6) & -473088  &  {\bf 1580672} & {\bf 671744}   & 
{\bf 101376}  &  
{\bf 4352} 
&  {-16} \st\\[1ex] \hline
(3,6) &  - 779104 &  {\bf 15219528} & {\bf 7997655}  & 
{\bf 1738664}
&  {\bf 149226} & {\bf 2164}  \st\\[1ex] \hline
 \hline 
\end{tabular}
\caption{Some results for $-B_6$ in the $\ZZZ_2$ CHL model 
for different values of $Q^2$, $P^2$ and
$Q . P$. The boldfaced entries are for charges
which satisfy the constraints \refb{egam2n}. We have only
given the results for $2 Q^2 \le P^2$, since due to a symmetry of
$\wt\Phi$ the $B_6$ index has a symmetry under $P^2\leftrightarrow
2 Q^2$\cite{0605210}.} \label{t2}
\end{center} }
\end{table}

\begin{table} {\small
\begin{center}\def\st{\vrule height 3ex width 0ex}
\begin{tabular}{|l|l|l|l|l|l|l|l|l|l|l|} \hline 
$(Q^2,P^2)\backslash Q.P$ &  -2 
& 0 & 1 & 2 & 3 & 4\st\\[1ex] \hline \hline
(2/3,2) & - 1458   &  {\bf 540 }
&  {\bf 27} &  0 & 0 & 0 \st\\[1ex] \hline
(2/3,4) & -  5616   & {\bf 3294}
&  {\bf 378 } & {0 } & 0 & 0 \st\\[1ex] \hline
(4/3,4) & -21496  &  {\bf 23008 } & {\bf  4912} &  {\bf 136 }
&  {0} & 0  \st\\[1ex] \hline
(2/3,6) & - 18900  &  {\bf 16200 }
&  {\bf  2646} & {54} & 0  & 0  \st\\[1ex] \hline
(4/3,6) & - 70524  &  {\bf 128706} & {\bf 37422}   & 
{\bf 2484}  &  
{ 6} 
&  {0} \st\\[1ex] \hline
(2,6) &  - 208584 & 
{\bf 820404} &  {\bf 318267} & {\bf 37818 }  
&  {\bf 801 } & { 0 }  \st\\[1ex] \hline
 \hline 
\end{tabular}
\caption{Some results for $-B_6$ in the $\ZZZ_3$ CHL model 
for different values of $Q^2$, $P^2$ and
$Q . P$. The boldfaced entries are for charges
which satisfy the constraints \refb{econd}. We have only
given the results for $3 Q^2 \le P^2$, since due to a symmetry of
$\wt\Phi$ the $B_6$ index has a symmetry under $P^2\leftrightarrow
3 Q^2$\cite{0605210}.} \label{t3}
\end{center} }
\end{table}

\begin{table} {\small
\begin{center}\def\st{\vrule height 3ex width 0ex}
\begin{tabular}{|l|l|l|l|l|l|l|l|l|l|l|} \hline 
$(Q^2,P^2)\backslash Q.P$ &  -2 
& 0 & 1 & 2 & 3 & 4\st\\[1ex] \hline \hline
(2/5,2) & -   392  &  {\bf 100  }
&  {  1 } &  0 & 0 & 0 \st\\[1ex] \hline
(2/5,4) & -   1120   & {\bf 460 }
&  {\bf 20} & {0 } & 0 & 0 \st\\[1ex] \hline
(4/5,4) & -  3200 &  {\bf 2280  } & {\bf  240 } &  {0  }
&  {0} & 0  \st\\[1ex] \hline
(2/5,6) & -  2940  &  {\bf 1720  }
&  {\bf  125 } & { 0} & 0  & 0  \st\\[1ex] \hline
(4/5,6) & -  8380  &  {\bf 9180 } & {\bf  1460}   & 
{\bf  20}  &  
{  0} 
&  {0} \st\\[1ex] \hline
(6/5,6) &  -  21660 & 
{\bf  39960} &  {\bf  9345} & {\bf 390 }  
&  {0  } & { 0 }  \st\\[1ex] \hline
 \hline 
\end{tabular}
\caption{Some results for $-B_6$ in the $\ZZZ_5$ CHL model 
for different values of $Q^2$, $P^2$ and
$Q . P$. The boldfaced entries are for charges
which satisfy the constraints \refb{enes}. We have only
given the results for $5 Q^2 \le P^2$, 
since due to a symmetry of
$\wt\Phi$ the $B_6$ index has a symmetry under $P^2\leftrightarrow
5 Q^2$\cite{0605210}.} \label{t5}
\end{center} }
\end{table}

\begin{table} {\small
\begin{center}\def\st{\vrule height 3ex width 0ex}
\begin{tabular}{|l|l|l|l|l|l|l|l|l|l|l|} \hline 
$(Q^2,P^2)\backslash Q.P$ &  -2 
& 0 & 1 & 2 & 3 & 4\st\\[1ex] \hline \hline
(2/7,2) & -   162  &  {\bf 36  }
&  {  0 } &  0 & 0 & 0 \st\\[1ex] \hline
(2/7,4) & -   396   & {\bf 138 }
&  {\bf 3} & {0 } & 0 & 0 \st\\[1ex] \hline
(4/7,4) & -  968 &  {\bf 564  } & {\bf  40 } &  {0  }
&  {0} & 0  \st\\[1ex] \hline
(2/7,6) & -  918  &  {\bf 444  }
&  {\bf  18 } & { 0} & 0  & 0  \st\\[1ex] \hline
(4/7,6) & -  2244  &  {\bf 1916 } & {\bf  210}   & 
{0}  &  
{  0} 
&  {0} \st\\[1ex] \hline
(6/7,6) &  -  5184 & 
{\bf  6892} &  {\bf  1152} & {\bf 18 }  
&  {0  } & { 0 }  \st\\[1ex] \hline
 \hline 
\end{tabular}
\caption{Some results for $-B_6$ in the $\ZZZ_7$ CHL model 
for different values of $Q^2$, $P^2$ and
$Q . P$. The boldfaced entries are for charges
which satisfy the constraints \refb{enes}. We have only
given the results for $7 Q^2 \le P^2$, 
since due to a symmetry of
$\wt\Phi$ the $B_6$ index has a symmetry under $P^2\leftrightarrow
7 Q^2$\cite{0605210}.} \label{t7}
\end{center} }
\end{table}

\sectiono{Test of positivity of the index} \label{sfour}

As already mentioned, the negativity of $B_6$
has been proved explicitly in the limit when all the charges
become large keeping the ratios $Q . P/P^2$, 
$Q^2/P^2$ fixed\cite{0708.1270}, 
not only for heterotic string theory on
$T^6$ but all $\NN=4$ supersymmetric string theories
where the answer for $B_6$ is known. In this section we shall
try to test this for finite charges. Note that due to
the $(-1)^{Q.P}$ factor in \refb{egg1int}, negativity
of $B_6$ means positive (negative) sign for the Fourier
coefficients of $1/\wt\Phi$ for odd (even) powers
of $e^{2\pi i v}$.

We begin
with heterotic string theory on $T^6$. The results for
$-B_6$ in $\RR$ for a range of values of $Q^2$, $P^2$
and $Q . P$ have been shown in table \ref{t1}.
Clearly the entries have positive and negative values.
But for charges which satisfy the restrictions
given in \refb{eb6neg} we have represented the 
entries by bold faced letters, and as we can see, all the bold
faced entries are manifestly positive. 
We have in fact checked that 
up to all values of $Q^2$ and $P^2$
up to 10 and all values of $Q . P$,  the positivity of
$-B_6$ inside $\RR$ holds whenever \refb{eb6neg}
holds.

Similar analysis is possible for $\ZZZ_N$ CHL models. We have 
checked the positivity of $-B_6$ for several charges in these
models and the result is again in accordance with the general
prediction from the black hole side. Some of the 
results are shown in
tables \ref{t2}, \ref{t3}, \ref{t5} and \ref{t7}. 
We have in fact tested the required positivity of $-B_6$
for all values of $NQ^2, P^2\le 10$ and all allowed values
of $Q.P$.
We have not gone to
very high values of the charges, but
it is more important to test this
for low charges since we already know that the prediction
holds in the large charge limit.

In all the tables we have specifically displayed the results for
$Q\cdot P=-2$ sector to emphasize the need for focussing on
single centered black holes for the positivity test of $-B_6$.
Due to a $v\to -v$ symmetry of $\wt\Phi$ the index for negative
$Q.P$ values in the chamber $\RR$ can be related to the index
for positive $Q.P$ values in the chamber $\LL$ lying to the left of
the wall from 0 to $i\infty$. Thus the results for $Q.P=-2$ 
given in the tables can be
reinterpreted as the results for $Q.P=2$ in the chamber $\LL$,
and
the difference between $Q.P=-2$ and the $Q.P=2$ entries
in the tables can be accounted for by the wall crossing
formula 
across the wall connecting 0 to $i\infty$. 
As we move from $\RR$
to $\LL$ across this wall new two centered configurations of a 
pair of half-BPS states, carrying charges $(Q,0)$ and $(0,P)$, 
appear. As can be seen from the tables, 
the negative contribution to
$-B_6$ from these states overwhelm the positive contribution
from single centered black holes for low values of the charges.

{\bf Acknowledgement:}  
I would like to thank 
Atish
Dabholkar, Joao Gomes  and
Sameer Murthy for useful discussions. 
I would also like to thank the Simons Center at Stony Brook for
hospitality and the participants at the Simons workshop for
useful discussions during the course of this work.
This work 
was supported in part by the J. C. Bose fellowship of the Department of
Science and Technology, India, the DAE
project 11-R\&D-HRI-5.02-0304, and by
the Chaires Internationales de Recherche Blaise Pascal, France.

\end{document}